\documentclass[useAMS,usenatbib]{mn2e} 

\usepackage{graphicx}
\usepackage{aas_macros} 
\usepackage{soul} 
\usepackage{bm} 

\input epsf

\usepackage{suffix}

\renewcommand{\em}{\it}  

\newcommand\LCDM{$\Lambda$CDM~}  
\newcommand\LCDMp{$\Lambda$CDM}  
\newcommand\LCDMeq{\Lambda{\rm CDM}}  

\newcommand{\Mpc}{\mbox{ Mpc}}

\newcommand{\ten}[1]{\times 10^{#1}}

\newcommand{\dint}[2]{{\rm d}^{#1}#2\;}

\newcommand{\mean}[1]{\left\langle{#1}\right\rangle} 
\newcommand{\partder}[2]{\frac{\partial #1}{\partial #2}} 
\newcommand{\fullder}[2]{\frac{{\rm d}#1}{{\rm d}#2}} 

\newcommand{\eeqp}{\;.\end{equation}}
\newcommand{\eeqc}{\;,\end{equation}}

\newcommand\labeq[1]{\label{eq:#1}}
\newcommand\refeq[1]{Eq.~(\ref{eq:#1})}
\WithSuffix\newcommand\refeq*[1]{(\ref{eq:#1})}

\newcommand\reffig[1]{Figure \ref{fig:#1}}

\newcommand\labsec[1]{\label{sec:#1}}
\newcommand\refsec[1]{\S \ref{sec:#1}}

\newcommand{\sfig}[2]{\includegraphics[width=#2]{#1}}

\newcommand{\vctr}[1]{{\bm #1}} 

\newcommand{\vr}{\vctr{r}}
\newcommand{\vs}{\vctr{s}}
\newcommand{\vk}{\vctr{k}}

\newcommand{\skypos}{\bm{\hat\theta}}

\newcommand{\dtempISW}{\frac{\Delta T^{\rm ISW}}{T}}
\newcommand{\dtempCC}{\frac{\Delta T^{\rm CC}}{T}}
\newcommand{\rmax}{r_{\rm max}}

\newcommand{\aCC}{a^{\rm CC}}

\newcommand{\Covar}{{\rm Cov}}
\newcommand{\Cov}{\mathcal{C}}

\newcommand{\gal}{{\rm gal}}
\newcommand{\rhogal}{\rho^\gal}
\newcommand{\rhogalbar}{\overline\rho^{\,\gal}}
\newcommand{\drhogal}{D}

\newcommand{\joint}{\Delta}

\newcommand{\moreeq}{\times\:}

\newcommand{\Cfig}[2]{
    \begin{figure}
    \begin{center}
    \sfig{#1.eps}{1.0\columnwidth}
    \caption{{\small #2}}
    \label{fig:#1}
    \end{center}
    \end{figure}
}

\title[Complementarity of ISW and RSD]
{The Complementarity of Redshift-space Distortions and the Integrated Sachs-Wolfe Effect: A 3D Spherical Analysis}
\author[C. Shapiro et al.\ ]{
C. Shapiro\thanks{Charles.A.Shapiro@jpl.nasa.gov}\footnotemark[0],
R.G. Crittenden,
W.J. Percival, \\
Institute of Cosmology \& Gravitation, University of Portsmouth, Dennis Sciama Bldg., Portsmouth, PO1 3FX, UK}

\begin{document}

\date{Version as of \today}

\pagerange{\pageref{firstpage}--\pageref{lastpage}} \pubyear{2011}

\maketitle

\label{firstpage}

\begin{abstract}
Assuming General Relativity is correct on large-scales, Redshift-Space Distortions (RSDs) and the Integrated Sachs-Wolfe effect (ISW) are both sensitive to the {\em time derivative} of the linear growth function.  We investigate the extent to which these probes provide complementary or redundant information when they are combined to constrain the evolution of the linear velocity power spectrum, often quantified by the function $f(z)\sigma_8(z)$, where $f$ is the logarithmic derivative of $\sigma_8$ with respect to $(1+z)$.  Using a spherical Fourier-Bessel (SFB) expansion for galaxy number counts and a spherical harmonic expansion for the CMB anisotropy, we compute the covariance matrices of the signals for a large galaxy redshift survey combined with a CMB survey like Planck.  The SFB basis allows accurate ISW estimates by avoiding the plane-parallel approximation, and it retains RSD information that is otherwise lost when projecting angular clustering onto redshift shells.  It also allows straightforward calculations of covariance with the CMB.  We find that the correlation between the ISW and RSD signals are low since the probes are sensitive to different modes. For our default surveys, on large scales ($k<0.05 \Mpc/h$), the ISW can improve constraints on $f\sigma_8$ by more than 10\% compared to using RSDs alone. In the future, when precision RSD measurements are available on smaller scales, the cosmological constraints from ISW measurements will not be competitive; however, they will remain a useful consistency test for possible systematic contamination and alternative models of gravity.
\end{abstract}


\section{Introduction} 

Large-scale structure formation is a valuable tool for discriminating among cosmological models.  When trying to distinguish \LCDM from its alternatives, one typically characterizes large-scale structure by $P(k,z)$, the matter power spectrum as a function of scale and redshift.  Various late Universe observations, such as galaxy clustering, galaxy cluster counting and weak gravitational lensing, are sensitive to this function, providing snapshots of the amount of structure at different epochs.
The dynamics of the Universe also produce observations that are sensitive to the {\em rate} of structure growth at a given epoch.  Such dynamical probes include redshift-space distortions (RSDs) and the Integrated Sachs-Wolfe effect (ISW).  These probes provide useful consistency checks on models of structure formation and can be used to distinguish models of dark energy from modifications to General Relativity (GR).

RSDs are an effect seen in galaxy redshift surveys due to the peculiar motions of galaxies.  In the absence of peculiar motions, the redshift of each galaxy corresponds to a unique distance $\vr$ from the observer; however, the line-of-sight component of a galaxy's peculiar velocity $\vctr{v}$ creates an additional Doppler shift that distorts the observed redshift, thereby ``moving'' the galaxy to a different inferred distance or {\em redshift-space} distance, $\vs(\vr,\vctr{v})$.  The net effect of peculiar velocities is to enhance the apparent galaxy clustering along the line-of-sight in redshift-space \citep{kaiser_1987,Fisher:1994fk}.  On scales larger than $\sim$10 Mpc, galaxies act as tracers of the bulk flow of matter and, by the continuity equation for a perfect fluid, the matter density at a particular position grows as matter flows in from surrounding regions.  It follows that peculiar velocities -- and therefore RSDs -- provide a measurement of the structure growth rate. 

At redshifts $z<0.3$, the measurements from the 2dF Galaxy
  Redshift Survey (2dFGRS; \citealt{colless03}) and the Sloan Digital
  Sky Survey (SDSS; \citealt{york00}) provide the best current
  constraints on RSD \citep{peacock01,hawkins03,percival04,pope04,zehavi05,okumura08,Cabre:2009uq,cabre09}. At higher redshifts, RSD have been
  measured in the VIMOS-VLT Deep Survey (VVDS)
  \citep{lefevre05,garilli08}, and Wiggle-Z \citep{drinkwater10}
  surveys \citep{guzzo09,Blake:2011fk}. These results are all fully
  consistent with the standard \LCDMp+GR model. 
RSDs also produce a signal when angular clustering measurements are made of projected data as RSD affect any redshift-dependent binning of galaxies \citep{Fisher:1994fk,nock10}. A correction for this has been included in previous analyses of angular clustering from photometric redshift surveys \citep{blake07,padmanabhan07}. There is a window of opportunity where surveys such as the Dark Energy Survey \citep[DES;][]{des_2005}, can provide cutting edge RSD constraints \citep{ross11}, before surveys such as BigBOSS \citep{schlegel09b} and Euclid \citep{Laureijs:2011uq} extend the redshift range of spectroscopic RSD measurements out to that of the photometric redshifts.

The ISW occurs when CMB photons traverse evolving potential wells.  If the gravitational potential around an overdensity decays (grows) as a photon crosses it, the photon will emerge with a net blueshift (redshift), which contributes to the overall CMB anisotropy.  In GR, potential wells in the late Universe are static during matter domination, reflecting a balance between structure growth and the background expansion of the Universe.  Once structure growth slows, due to e.g.~the presence of dark energy, potentials decay in absolute value.  The ISW is difficult to measure using the CMB alone but can be detected by cross-correlating the CMB with foreground objects that trace the large-scale structure \citep{Crittenden:1995ak}.  

Initial CMB-galaxy cross-correlation measurements using the COBE data failed to find a signal, but this changed dramatically with the arrival of the WMAP data \citep{bennett_halpern_etal_2003}.  Weak correlations, at the $2-3 \sigma$ level, were soon seen between the CMB and numerous probes of large scale structure, such as the NVSS radio galaxy survey and the X-ray background \citep{boughn_crittenden_2004,Nolta:2003uy}, and in optical surveys like the SDSS \citep{Fosalba:2003kx,Scranton:2003in, Fosalba:2003iy,Padmanabhan:2005uq,Giannantonio:2006du}, while weaker indications been seen using the shallower 2MASS infrared survey \citep{afshordi_loh_etal_2004,Rassat:2006kq,francis_peacock_2010}.
Combining these probes \citep{Giannantonio:2008zi,Ho:2008bz} raises the significance to the $4 \sigma$ level \citep[see also][for a comprehensive summary]{Dupe:2011fk}. 
Planck should not dramatically differ from WMAP on the large scales to which the ISW is most sensitive, but its greater resolution and frequency coverage will enable greater control of possible systematic contaminations from the galaxy and extra-galactic point sources.   Improvements to ISW measurements will ultimately require better large scale structure data covering most of the sky, such as that provided by radio surveys \citep[e.g.][]{raccanelli_zhao_etal_2011} or other frequencies \citep[e.g.~DES and WISE: ][]{des_2005,wright_eisenhardt_etal_2010}.

Large redshift surveys can be cross-correlated with the CMB to measure the ISW effect in parallel with providing galaxy clustering and RSD measurements.  Indeed, this combination has already been explored by \citet{Cabre:2009uq} using SDSS and WMAP data.  
Our goal in this paper is to understand the potential of a larger galaxy survey such as BOSS to constrain the growth rate of large-scale structure by combining the two datasets while accounting for the covariance between them.  We will work within the context of GR.
If one does not assume that GR holds, then RSD and ISW measurements  contain complementary information: allowing for the most general  metric-based models, ISW measurements depend on both time-like and  space-like metric fluctuations, while RSD only depend on time-like  fluctuations.  Once Einstein's equations are assumed, both probes can be shown to measure the amplitude of the peculiar velocities on large scales, $f(z)\sigma_8(z)$, where $\sigma_8(z)$ is the normalization of the linear matter power spectrum and 
\begin{equation}
f(z)\equiv-\fullder{\ln\sigma_8(z)}{\ln(1+z)}
\;.\end{equation}
When combining the ISW with RSDs, one naively expects the observables to be correlated  since they both arise from fluctuations in gravitational potentials,  but we will show that they are largely uncorrelated. 

  
For our analysis, we decompose the galaxy number density into discrete spherical Fourier-Bessel (SFB) modes (eigenfunctions of the Laplacian operator in spherical coordinates).  This basis provides a natural way to relate the 3D galaxy field to the 2D spherical harmonics of the CMB while avoiding approximations which may be unsuitable for either data set.  For example, the distant observer and plane parallel approximations (typically applied to galaxy counts) poorly describe the ISW, which is only relevant for very large distance scales and wide angles.  Projecting galaxies into redshift slices (typically done for ISW) discards radial information about galaxy positions, which weakens the RSD signal.  We could have chosen to bin very finely in redshift, but this would lead to a highly correlated and ill-conditioned covariance matrices.  \textbf{Previous works have also found the SFB basis useful for analysing various probes of large-scale-structure \citep{Fisher:1995mz,Schmoldt:1999ve,Heavens:2003ly,Castro:2005ys,Erdogdu:2006zr,Erdogdu:2006fr,Abramo:2010rt,Rassat:2011gf}.}

\subsection*{Outline}

We begin in \refsec{fiducial} with a brief summary of the survey assumptions and the cosmological model on which we base our subsequent calculations.  In \refsec{spharm}, we review the SFB expansion and apply it to the ISW signal and to the galaxy number density in redshift-space.  In \refsec{covar}, we compute the covariances and correlations of these signals.  In \refsec{forecast}, we forecast the ability of RSDs and the ISW to constrain the growth rate when combined.  We conclude in \refsec{conclusions}.  The appendices discuss issues related to the sampling of discrete $k$-modes in the SFB formalism.

\section{Description of Fiducial Survey} \labsec{fiducial}

We consider a spectroscopic redshift survey which measures the redshift $z$ and sky position $\skypos$ for a large sample of galaxies.  For simplicity, the survey is assumed to cover the full solid angle on the sky and have a mean observed comoving galaxy density $\rhogal(z)$ that is 
constant over the redshift range, $(0, z_{\max})$.  We assume that galaxy redshifts and densities are similar to those expected for the BOSS survey, and take $z_{\max}=0.6$ and $\rhogal(z)=\rhogalbar=3\ten{-4}\;h^3/\Mpc^3$ with $H_0=h\times100$ km/s/Mpc.  Our fiducial survey subsequently has 4 times the volume of BOSS, which will only survey 1/4 of the sky.
The galaxy sample is assumed to obey a scale-independent linear bias with respect to dark matter clustering:
\begin{equation}
\delta^{\gal}(k)=b(z)\delta(k)
\;,
\end{equation}
where we ignore the gauge-dependent effects of the bias definition in GR \citep[these introduce corrections for horizon-sized scales; see][]{challinor_lewis_2011,bonvin_durrer_2011}. We also assume constant galaxy clustering (CGC), i.e.~that the observed power spectrum of galaxy density $P^{\gal}(k)$ is constant in $z$.  Physically, CGC means that the galaxy bias is related to the linear growth rate by
\begin{equation} \labeq{CGC}
b(z) = \frac{b(0)}{G(z)}
\end{equation}
where matter density modes grow by $G(z)\equiv\delta(k;z)/\delta(k;z=0)$ in linear theory.  Linear bias and CGC are suitable approximations for BOSS, for which $b(0)=1.7$ \citep{schlegel09a}.
We restrict our analysis to galaxy density modes with $k\le 0.1 h/$Mpc; for our purposes, these modes are described sufficiently accurately by linear theory.

For Planck, we assume all-sky, cosmic-variance limited measurements of the CMB temperature anisotropy.  We restrict our analysis to large angular scales ($>$1 deg) for which details of the beam will be irrelevant.

Our fiducial cosmological model is a flat \LCDM model with the following parameters:
\begin{center}
  \begin{tabular}{@{} ccccc @{}}
    \hline
    $\Omega_m$ & $\Omega_b$ & $h$ & $n$ & $\sigma_8$  \\
    0.27 & 0.045 & 0.72 & 1 & 0.75  \\
    \hline
  \end{tabular}
\end{center}
The parameters $\Omega_m$ and $\Omega_b$ are the total matter and baryonic matter densities as fractions of the critical density; $n$ is the tilt of the primordial spectrum of scalar fluctuations; and $\sigma_8$ is the normalization of the  matter power spectrum today, expressed as the RMS of linear density fluctuations smoothed on scales of 8 Mpc/$h$.
All figures and calculations use these parameters unless otherwise noted.  
We work in the Newtonian gauge, for which the perturbed Friedman-Robertson-Walker metric is described by
\begin{equation}
ds^2 = a^2[\; -(1+2\psi)d\eta^2 + (1+2\phi) (dr^2 + r^2 d\Omega) \;]
\;,\end{equation}
where $\eta$ is the conformal time,  $\phi$ and $\psi$ are the gravitational potentials, and $a=(1+z)^{-1}$ is the cosmic scale factor.
We calculate the linear matter power spectrum (without baryon wiggles) using the fitting formulae of \citet{eisenstein_hu_1999}.
\section{The Spherical Fourier-Bessel Expansion applied to Large-Scale Structure} \labsec{spharm}

In this section, we use a SFB expansion to relate the matter power spectrum to both the 3D galaxy density field and the 2D CMB anisotropy.  This approach allows a straightforward calculation of covariance matrices for the two observables, as we show in \refsec{covar}.

\subsection{Review of the SFB expansion}

The SFB expansion allows us to express any scalar function in terms of the eigenfunctions of the Laplacian operator in spherical coordinates.  Galaxy number density can be expanded thusly:
\begin{equation}
\rhogal(\vr) = \sum_{l=0}^\infty \,\sum_{m=-l}^l \,\sum_n c_{ln} \rhogal_{lmn}j_l(k_{ln}r)Y_{lm}(\skypos)
\end{equation}
where $\vr=(r,\skypos)$ is the comoving distance vector, $j_l$ is a spherical Bessel function, $Y_{lm}$ is a spherical harmonic, and $c_{ln}$ is a normalization constant defined below.
For a given $l$, the $k_{ln}$ and $c_{ln}$ will depend on a choice of boundary conditions.  \citet{heavens_taylor_1995} proposed using the discrete wavenumbers that extremize the spherical Bessel functions $j_l(k_{ln}r)$ on the survey boundary (Neumann boundary conditions):
\begin{equation} \labeq{boundary}
\left. \fullder{}{r} j_l(k_{ln}r) \right|_{r_{\max}} = 0
\;,\end{equation}
where $r_{\max}\equiv r(z_{\max})$.  This choice avoids redshift-space distortions on the survey boundary\footnote{The boundary conditions restrict our ability to reconstruct certain real-space features of the density field; however, we are only interested in the power spectrum of modes that are sampled by the survey, not accurate reconstructions of the density and velocity fields.  See Appendix~\ref{sec:ksamp} for further discussion.}, i.e.~the survey boundary is a sphere of radius $r_{\max}$ in both real-space and redshift-space.  With this choice of boundary conditions, Sturm-Liouville theory guarantees that the $j_l(k_{ln}r)$ form a complete basis for the radial part of the expansion on the finite interval $(0,r_{\max})$ \citep{wang_ronneberger_etal_2008}.  

For our full-sky survey, the range of $l$ and $n$ required is determined by the radial distribution of galaxies
and the desired 3-D resolution.  The $l$ value
determines the wavenumber perpendicular to the line of sight, while the
$k_{ln}$ indicate the wavenumber along the line of sight.  For a fixed
3-D wavenumber, $k_{\max}$, higher $l$ values have fewer line of sight
modes.
For each $l \ge 2$, we order the discrete wavenumbers such that $k_{ln}$ monotonically increases with $n$, and we define the smallest $k_{ln}$ to have $n=1$.  For $l=2$, we find that we need all $n$ up to $n\approx r_{\max}k_{\max}/\pi$ to account for all $k_{2n}<k_{\max}$.  Larger $l$ require fewer $n$, and the largest included $l$ may have only one mode ($n=1$) below the maximum $k$.  In our case, in order to include all $k_{l1}<k_{\max}$, we find that we need $l$ up to $l\approx r_{\max}k_{\max}$.  The smallest $k$ (longest wavelength) included in the analysis is $k_{21}=3.342/r_{\max}$; for our fiducial survey with $z_{\max}=0.6$ and $r_{\max}=1563 \Mpc/h$, we have $k_{21}=0.00214\,h$/Mpc. 

The spherical Bessel functions obey the following orthogonality relation:
\begin{equation} \labeq{norm_const}
\int_0^{r_{\max}}\!\!\!\!\!\!\dint{}{r} j_l(k_{ln}r)j_l(k_{ln'}r) r^2 = c_{ln}^{-2} \delta^K_{nn'}
\end{equation}
where $\delta^K$ is the Kronecker delta, and
\begin{equation}
 c_{ln}^{-2} = \frac{r_{\max}^3}{2}\left[1-\frac{l(l+1)}{(k_{ln}r_{\max})^2}\right] j_l^2(k_{ln}r_{\max})
\end{equation}
The spherical harmonics are also orthogonal:
\begin{equation}
\int_{4\pi}\dint{2}{\theta}Y_{lm}(\skypos)Y^*_{l'm'}(\skypos) = \delta^K_{ll'}\delta^K_{mm'}
\;.\end{equation}
Therefore, the SFB coefficients are given by
\begin{equation}
\rhogal_{lmn} = c_{ln} \int_{4\pi}\int_0^{r_{\max}}\!\!\!\!\!\!\dint{3}{r} \rhogal(\vr) j_l(k_{ln}r) Y^*_{lm}(\skypos)
\;.\end{equation}

\subsection{Galaxy Density Fluctuations in Redshift-Space}


In a flat Friedmann-Robertson-Walker (FRW) metric, redshift and comoving distance are related by $\fullder{z}{r}=H(z)$, where $H(z)=\dot a/a$ is the Hubble parameter.  This relation is applied to measured redshifts to obtain a redshift-distance, $\vs$, which includes distortions from peculiar velocities.  \citet[][hereafter P04]{percival04} applied the work of \citet[][hereafter HT95]{heavens_taylor_1995} to the 2dFGRS, providing details of a practical implementation of the SFB analysis to galaxy counts. Here, we reproduce the calculation described by P04, simplified for an all-sky survey with CGC, no weighting, and no non-linear RSDs on small scales (Fingers-of-God).  In the notation of P04, Greek subscripts refer to a triplet of indices $\nu\equiv(l_\nu m_\nu n_\nu)$ and $j_\nu(r)\equiv j_{l_\nu}(k_{l_\nu n_\nu}r)$.   

The galaxy overdensity in redshift space, $\drhogal(\vs)\equiv\rhogal(\vs)-\rhogalbar(\vs)$, can be transformed into its SFB coefficients,
\begin{equation}
\drhogal^s_{\nu} \equiv c_{\nu} \int_{4\pi}\int_0^{s_{\max}}\!\!\!\!\!\!\dint{3}{s} \drhogal(\vs) j_{\nu}(s) Y^*_{\nu}(\skypos)
\;, \end{equation}
and expressed as linear combinations of the real-space matter overdensity coefficients, $\delta_\mu$, at $z=0$:
\begin{equation} \labeq{drhogal}
{\drhogal}^s_\nu = \sum_\mu [ b(0)\,\Phi_{\nu\mu} + f(0) V_{\nu\mu} ] \delta_\mu 
\;.\end{equation}
There is a usual clustering term
\begin{eqnarray}
\Phi_{\nu\mu} & \equiv & \delta^K_{l_{\mu}l_{\nu}}\delta^K_{m_{\mu}m_{\nu}} c_\nu c_\mu  \nonumber \\
	&  & \moreeq \int_0^{\rmax}\!\!\!\!\!\! \dint{}{r}r^2 j_\nu(r)j_\mu(r) \rhogalbar(r) G(z)\frac{b(z)}{b(0)}
\end{eqnarray}
and a RSD term
\begin{eqnarray}
V_{\nu\mu} & \equiv & \delta^K_{l_{\mu}l_{\nu}}\delta^K_{m_{\mu}m_{\nu}} \frac{c_\nu c_\mu}{k_\mu^2} \nonumber \\
    &  & \moreeq \int_0^{\rmax}\!\!\!\!\!\! \dint{}{r}r^2 j'_\nu(r) j'_\mu(r) \rhogalbar(r) G(z)\frac{f(z)}{f(0)}
\end{eqnarray}
where $z=z(r)$ is evaluated on our past light-cone and $j' = dj/dr$.  Note that for an all-sky survey with no masking, $\Phi_{\mu\nu}$ is symmetric but $V_{\nu\mu}$ is not: $f(0)V_{\nu\mu}$ is the contribution to the observed galaxy density mode ${\drhogal}^s_\nu$ from velocities sourced by matter mode $\delta_\mu$.  
We will use the approximation of \citet{Linder:2005in}, who showed that $f(z)\approx\Omega_m(z)^{0.55}$ for \LCDMp. \citep[other approximations have been derived by e.g.][]{peebles_1980,lahav_lilje_etal_1991,wang_steinhardt_1998}.

Since our simplified $\Phi$ and $V$ matrices are independent of $m$ and block-diagonal in $l$ and $m$, the ${\drhogal}^s_\mu$ are more simply expressed as
\begin{eqnarray}
{\drhogal}^s_{lmn} & = & \sum_{n'} [ b(0)\,\Phi^l_{nn'} + f(0)\,V^l_{nn'} ] \delta_{lmn'} \labeq{gal_mode_simple} \\
\Phi^l_{nn'} & \equiv &  c_{ln} c_{ln'} \int_0^{\rmax}\!\!\!\!\!\!  \dint{}{r}r^2 j_l(rk_{ln})j_l(rk_{ln'}) \nonumber \\
	  &  & \moreeq  \rhogalbar(r) G(z)\frac{b(z)}{b(0)} \labeq{Phi_lnn} \\
V^l_{nn'} & \equiv & \frac{c_{ln} c_{ln'}}{k_{ln'}^2} \int_0^{\rmax}\!\!\!\!\!\! \dint{}{r}r^2 j'_l(rk_{ln})j'_l(rk_{ln'}) \nonumber \\
	  & & \moreeq \rhogalbar(r) G(z)\frac{f(z)}{f(0)} \labeq{V_lnn}
\,.\end{eqnarray}
Using \refeq{norm_const}, the CGC condition of \refeq{CGC}, and the orthogonality of the spherical Bessel functions, we find that $\Phi^l_{nn'}$ is simply proportional to the identity matrix when $\rhogalbar(z)$ is constant:
\begin{eqnarray}
\Phi^l_{nn'} & = & c_{ln} c_{ln'} \rhogalbar  \int_0^{\rmax} \!\!\!\!\!\! \dint{}{r}r^2 j_l(rk_{ln})j_l(rk_{ln'}) \nonumber \\
 & = &  c_{ln} c_{ln'} \rhogalbar [c_{ln}^{-2}\delta^K_{nn'}] \; = \; \rhogalbar \delta^K_{nn'}
. \end{eqnarray} 

Sections of the RSD matrix $V^l_{nn'}$ are illustrated by solid lines in \reffig{matrix_sections} (assuming CGC).  Even in our idealized survey, the matrices exhibit significant mode mixing at low $l$.  Mathematically, the negative off-diagonal elements of $V^l_{nn'}$ arise because, unlike the spherical Bessel functions in \refeq{gal_mode_simple}, the Bessel function derivatives in \refeq{V_lnn} do not form a perfectly orthogonal basis.  This result contrasts with the familiar Kaiser result \citep{kaiser_1987}, which has no mode mixing:
\begin{equation}
{\drhogal}^s(\vk) = \delta(\vk)\left[b+(\hat\vk\cdot\hat\vr)^2 f \right]
\;.\end{equation}
Kaiser approximates the radial distortions as being along a particular Cartesian axis, which allows a plane-wave expansion.  Whereas the derivative of a single plane-wave is another plane-wave with the same frequency, spherical Bessel functions do not enjoy this property.  Therefore radial mode-mixing is an inevitable geometric effect of RSDs on large angular scales \citep{zaroubi_hoffman_1996,heavens_taylor_1995}. For the correlation function and its Legendre multipole expansions, the full wide angle effects including this mode mixing can be calculated using a tri-polar spherical harmonics expansion as described by \citet{szalay_matsubara_etal_1998,szapudi_2004,papai_szapudi_2008}, and tested by \citet{raccanelli_samushia_etal_2010}. 
Although we restrict our analysis to $k<0.1 h/$Mpc, the observable redshift-space modes, ${\drhogal}^s_{lmn}$, with $k_{ln}$ just below this cutoff will have contributions from real-space modes, ${\delta}_{lmn'}$, with $k_{ln'}$ just above the cutoff.  Therefore, computations must extend beyond the cutoff scale to ensure convergence.  We find that including modes up to $k=0.15 h/$Mpc is more than sufficient for our purposes and that we can safely ignore mixing between modes with $\Delta k>0.05 h$/Mpc; however, surveys with a non-trivial $\rhogalbar(z)$ or selection function will exhibit greater mode-mixing.  HT95 describe how to attenuate mode-mixing by optimally weighting the data.

\subsection{The Integrated Sachs-Wolfe Effect}

\Cfig{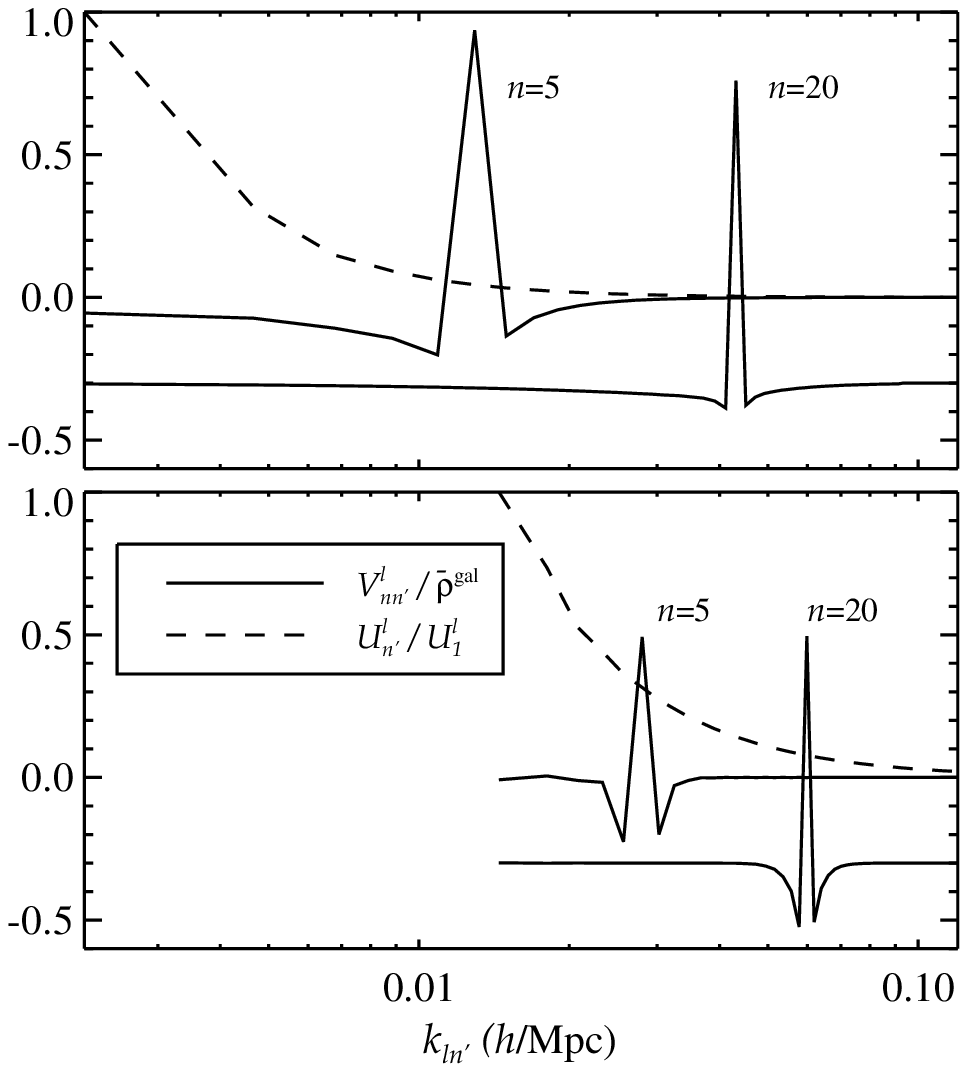}{Sections of the $V^l_{nn'}$ and $U^l_{n'}$ matrices for $l=2$ (top) and $l=20$ (bottom); respectively, these give the contributions to the RSD and ISW signals from a real-space matter density mode, $\delta_{lmn'}$.  For clarity, $V^l_{nn'}$ for $n$=20 has been offset by -0.3, and all matrices have been scaled to fit on a common vertical axis (see legend).  The central spikes of the $V^l_{nn'}$ occur where $n'=n$. Note that for a given $l$, the lowest available wavenumber is $k_{l1}$.}

The ISW contribution to the 2D CMB temperature anisotropy is \citep{sachs_wolfe_1967}
\begin{eqnarray}
\dtempISW(\skypos) &=& \int_{\eta_0}^{\eta_{\rm LS}} \dint{}{\eta} e^{-\tau(\eta)} \partder{ }{\eta}\left[\psi(\vr;\eta)-\phi(\vr;\eta)\right] \nonumber \\
&=& 2\int_{\eta_0}^{\eta_{\rm LS}} \dint{}{\eta} e^{-\tau(\eta)} \partder{\phi(\vr;\eta)}{\eta}
\end{eqnarray}
where $\eta$ is conformal time, $\eta_0$ is the present time, $\eta_{\rm LS}$ is the time at the last scattering surface.  
The last equality follows from assuming GR and the absence of anisotropic stress, so that $\psi=-\phi$.
The integration is done along a line-of-sight trajectory $\vr = (r,\skypos)$ where $r=r(\eta)$ is the comoving distance to an observed CMB photon at time $\eta$.   Although the integral is along our past lightcone, note that the time derivative $\partder{\phi}{\eta}$ is simply taken with respect to the global time coordinate, $\eta$.  The CMB scattering function $\tau(\eta)$ is much less than unity, so we will neglect the factor $e^{-\tau}$ hereafter.  Thus, we take 
\begin{equation}
\dtempISW(\skypos) = 2\int_{\eta_0}^{\eta_{\rm LS}}\!\!\!\! \dint{}{\eta} \partder{\phi(\vr;\eta)}{\eta} = 2\int_{0}^{r_{\rm LS}}\!\!\!\! \dint{}{r} \partder{\phi(\vr;z)}{r}
\;,\end{equation}
where we have changed variables to match the previous section.  The $z$ argument above reminds us that $\phi$ is being evaluated along our past light-cone at $z=z(r)$.

Although the ISW arises from all potentials between the observer and the last-scattering surface, surveys are primarily sensitive to potentials within the survey boundary, which is a sphere of radius $r_{\max}=r(\eta_{\max})$.  Therefore we write
\begin{equation}
\dtempISW(\skypos) = 2\int_{0}^{r_{\max}}\!\!\!\!\!\! \dint{}{r} \partder{\phi(\vr;z)}{r}  + 2\int_{r_{\max}}^{r_{\rm LS}} \!\!\!\! \dint{}{r} \partder{\phi(\vr;z)}{r}
\;,\end{equation}
and ignore the second term on the right since it has only a relatively small correlation with the galaxies in the survey.  We label the remaining term with a ``CC'' for ``cross-correlate'':
\begin{equation}
\dtempCC(\skypos) \equiv 2\int_{0}^{r_{\max}}\!\!\!\!\!\! \dint{}{r} \partder{\phi(\vr;z)}{r}
\;.\end{equation}
Expanding the potential into SFB modes, we find
\begin{eqnarray}
\dtempCC(\skypos) & \!\!\!\!= & \!\!\!2\int_{0}^{r_{\max}}\!\!\!\!\!\! \dint{}{r}
	\partder{}{r}\left[\, \sum_{lmn} c_{ln} \phi_{lmn}(z) j_l(k_{ln}r)Y_{lm}(\skypos) \right] \nonumber \\
  & \!\!\!\!= & \!\!\!2 \sum_{lmn} c_{ln} Y_{lm}(\skypos) \int_{0}^{r_{\max}}\!\!\!\!\!\!\!\! \dint{}{r}
	 j_l(k_{ln}r) \partder{}{r}\phi_{lmn}(z) \labeq{temp2pot}
\end{eqnarray}
where the $k_{ln}$ and $c_{ln}$ are (conveniently) the same as those for the galaxy survey.  The $z$ argument remains in $\phi_{lmn}(z)$ to account for linear growth.

Einstein's equations allow us to convert potential to density using
Poisson's equation,\footnote{Strictly speaking this equation depends on the gauge where the density
fluctuations are defined, but on sufficiently small scales all gauges
are equivalent.  For the bulk of the ISW signal-to-noise, this is a good
approximation, but the low $l$ results can be affected \citep{yoo_2009,yoo_fitzpatrick_etal_2009}. }
\begin{equation}
\nabla^2 \phi(\vr) = -4\pi G \delta\rho(\vr)
\;,\end{equation}
which simplifies substantially in the SFB basis:
\begin{equation}
\phi_{lmn}(z) = \frac{a^2}{k_{ln}^2} 4\pi G\; \delta\rho_{lmn}(z)
\;.\end{equation}
The scale factor $a=1/(1+z)$ accounts for the metric in an expanding Universe \cite[see e.g.][]{dodelson_2003}.  We can rewrite this as
\begin{equation} \labeq{pot2density_sph}
\phi_{lmn}(z) = \frac{3}{2}\Omega_m H_0^2 \frac{\delta_{lmn}}{k_{ln}^2} \frac{G(z)}{a}
\;,\end{equation}
where $\delta$ is the present comoving matter overdensity and $G(z)$ is the growth factor (not to be confused with the gravitational constant, which been absorbed into $H_0^2$).  Plugging \refeq{pot2density_sph} into \refeq{temp2pot} and defining
\begin{equation}
\aCC_{lm} \equiv \int \dint{2}{\theta} \dtempCC(\skypos) Y^*_{lm}(\skypos)
\;,\end{equation}
we find that it is a linear combination of the $\delta_{lmn}$:
\begin{equation}
\aCC_{lm} = 3 \Omega_m H_0^2 \sum_n \frac{c_{ln}}{k_{ln}^2}\delta_{lmn}
		\int_0^{r_{\max}}\!\!\!\!\!\! \dint{}{r} j_l(k_{ln}r) \fullder{}{r}\left[\frac{G(z)}{a} \right]
.\end{equation}
To reiterate, $\aCC_{lm}$ is not the full ISW contribution to the CMB anisotropy -- it's just the contribution from density modes in our analysis, which are restricted by the galaxy survey geometry and our boundary conditions.

We now rewrite $\aCC_{lm}$ in terms of $f(z)$:
\begin{eqnarray}
\aCC_{lm} &=& 3 \Omega_m H_0^2 \sum_n \frac{c_{ln}}{k_{ln}^2}\delta_{lmn} \nonumber \\
		& & \moreeq \int_0^{r_{\max}}\!\!\!\!\!\! \dint{}{r} j_l(k_{ln}r) H(z)\fullder{}{z}\left[(1+z)G(z) \right] \nonumber \\				
&=& 3 \Omega_m H_0^2 \sum_n \frac{c_{ln}}{k_{ln}^2}\delta_{lmn} \nonumber \\
& & \moreeq	\int_0^{r_{\max}}\!\!\!\!\!\! \dint{}{r} j_l(k_{ln}r) H(z)G(z) [1-f(z)]
\end{eqnarray}
where we have used $f=-\fullder{\ln G}{\ln (1+z)}$ and $\fullder{z}{r}=H(z)$.  For convenience, we define
\begin{equation} \labeq{U_ln}
U^l_n \equiv 3\Omega_m H_0^2 \frac{c_{ln}}{k_{ln}^2} \int_0^{r_{\max}}\!\!\!\!\!\! \dint{}{r} j_l(k_{ln}r) H(z)G(z) [1-f(z)]
\end{equation}
so that
\begin{equation} \labeq{aISW_simple}
\aCC_{lm} = \sum_n U^l_n\delta_{lmn}
\;.\end{equation}
Sections of the $U^l_n$ matrix are shown in \reffig{matrix_sections}.

\subsection{Comparison of Sensitivity to Large-Scale Structure Growth}\labsec{growth-compare}

\Cfig{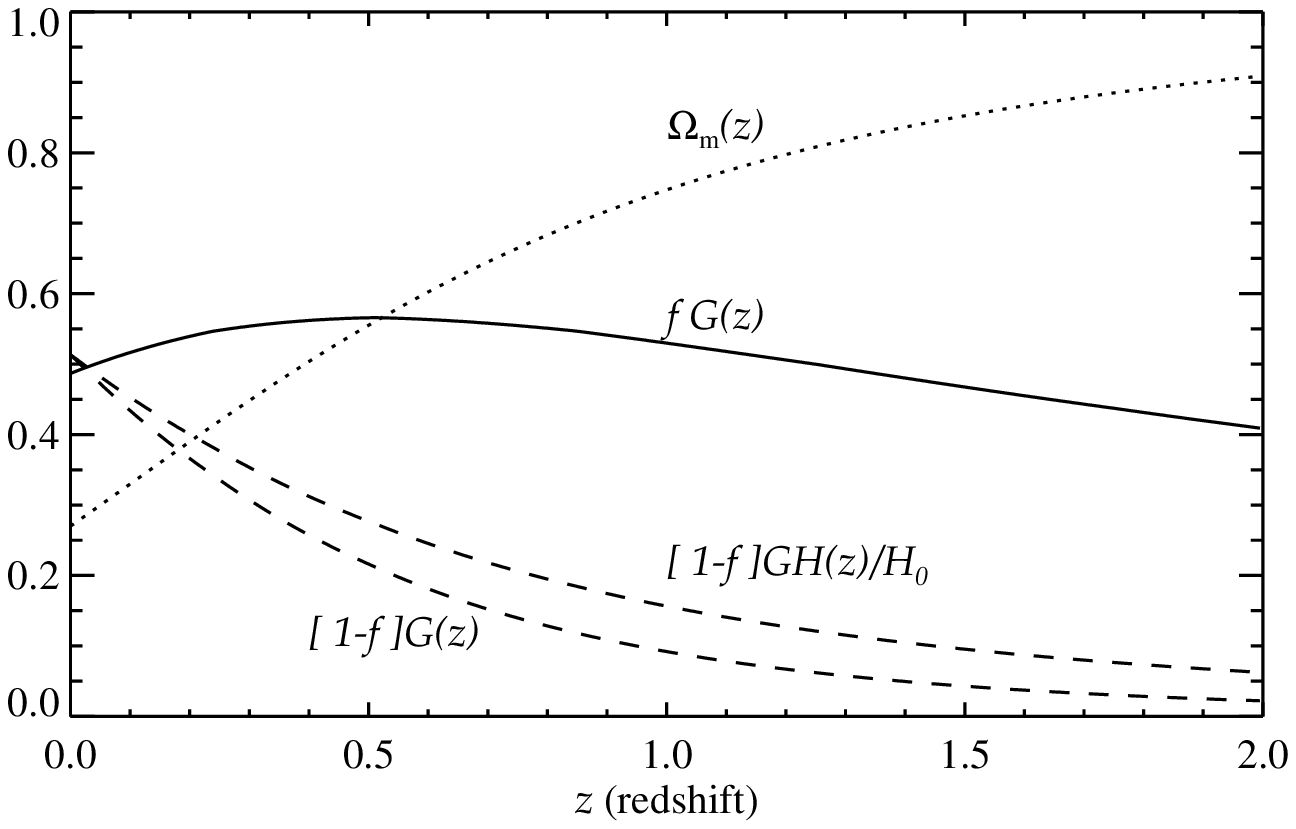}{ Functions related to the linear growth of large-scale structure.  RSDs are sensitive to $f(z)G(z)$ while the ISW is sensitive to $[1-f(z)]G(z)$. }

Notice by \refeq{V_lnn} and \refeq{U_ln} that RSDs and the ISW are sensitive to complementary functions of the linear growth and growth rate: $fG(z)$ and $(1-f)G(z)$, respectively.  Both functions can be expressed in terms of $f\sigma_8(z)$, since
\begin{equation}
\sigma_8(z)=\sigma_8(0)G(z)
\;.\end{equation}
There is a relative $H(z)$ factor between the RSD and ISW dependences; however, these data will not be a primary probe of that function.  As shown by the dashed curves in \reffig{growfuncs}, the ISW signal is strongest at late times when the Universe is becoming $\Lambda$ dominated and gravitational potentials are decaying most rapidly.  The RSD signal is strongest at intermediate redshifts but maintains a relatively constant signal over the range plotted.  This reflects a balance between $f(z)$, which drops as the Universe leaves matter domination, and $G(z)$, which continues to increase.   The real-space galaxy clustering, described by \refeq{Phi_lnn}, is sensitive to $b\sigma_8(z)$, which is constant in the case of CGC.

Since both the RSD and ISW signals can be expressed as sums over the matter density modes, $\delta_{lmn}$, we generally expect some correlation between them.  As we show in the next section, the degree of correlation is determined by a contraction of the $U^l_n$ and $V^l_{nn'}$ matrices.  Due to their relatively orthogonal shapes (shown in \reffig{matrix_sections}), the correlation turns out to be low, which makes the ISW and RSDs independent probes of the structure growth rate despite being based on closely related underlying physics.

\section{Covariances and Correlations}  \labsec{covar}

Having written down expressions for the redshift-space galaxy density modes ${\drhogal}^s_{lmn}$ and the ISW contribution to the CMB anisotropy $\aCC_{lm}$, we will now calculate their covariances -- measurable quantities that we can predict.  
Since ${\drhogal}^s_{lmn}$ and $\aCC_{lm}$ are both linear combinations of the real-space matter density modes, $\delta_{lmn}$, the covariance between any two observable modes will be a linear combination of $\mean{\delta_{lmn}\delta^*_{l'm'n'}}$, where angle-brackets denote an ensemble average.  Fortunately, in linear theory,
\begin{equation} \labeq{pk_sph}
\mean{\delta_{lmn}\delta^*_{l'm'n'}} = P(k_{ln})\delta^K_{ll'}\delta^K_{mm'}\delta^K_{nn'}
\;.\end{equation}
Hence, all covariances will simplify substantially and be expressible in terms of $P(k)$.  By the symmetry of the $\Phi$, $V$, and $U$ matrices, the covariances between observable modes will vanish unless $l=l'$ and $m=m'$, so we can restrict our attention to these cases.

Using \refeq{gal_mode_simple} and \refeq{pk_sph}, the covariances of the density modes are
\begin{eqnarray}
\mean{{\drhogal}^s_{lmn}{\drhogal}^{s*}_{lmn'}} &=& \sum_{n''} P(k_{ln''})[b(0)\,\Phi^l_{nn''}+f(0)\,V^l_{nn''}] \nonumber \\
& & \moreeq [b(0)\,\Phi^l_{n'n''}+f(0)\,V^l_{n'n''}]
\end{eqnarray}
Note that the order of the subscripts of $V^l_{nn'}$ is important since that matrix is asymmetric.  The measured modes, $\hat{\drhogal}^s_{lmn}$, will include shot noise; for a constant-density/all-sky survey, the total covariance will be
\begin{equation}
\mean{\hat{\drhogal}^s_{lmn}\hat{\drhogal}^{s*}_{lmn'}} = \mean{{\drhogal}^s_{lmn}{\drhogal}^{s*}_{lmn'}} + \Lambda^l_{nn'}
\;.\end{equation}
In our simplified case of an all-sky survey with no weighting or masking, the shot noise term is given by $\Lambda^l_{nn'}=\Phi^l_{nn'}$.

We cannot measure the ISW signal, $\aCC_{lm}$, in isolation: we must correlate the galaxy density with the total measured CMB anisotropy, which includes contributions from the early Universe and detector noise.  That is, we can only measure $\mean{\hat a_{lm}\hat{\drhogal}^{s*}_{lmn}}$.
However, the predominant galaxy-CMB correlations are sourced by gravitational potentials within the survey boundary, so we can safely ignore  CMB anisotropies originating at higher $z$ and on the surface of last-scattering (chance correlations with these sources will be accounted for in our covariance matrix).  Furthermore, we do not expect correlations between the noise of the two types of observables.  Hence, we let $\mean{\hat a_{lm}\hat{\drhogal}^{s*}_{lmn}}=\mean{\aCC_{lm}{\drhogal}^{s*}_{lmn}}$.  Combining \refeq{aISW_simple}, \refeq{gal_mode_simple}, and \refeq{pk_sph}, we find that
\begin{eqnarray}
\mean{\aCC_{lm}{\drhogal}^{s*}_{lmn}} = \sum_{n'} P(k_{ln'}) [b(0)\,\Phi^l_{nn'}+f(0)\,V^l_{nn'}]U^l_{n'} \labeq{CMB-gal-covar}
\;.\end{eqnarray}
Note that the $VU$ term is the ``effect of RSDs on the ISW signal,'' which we discuss below.
The covariances of the $\aCC_{lm}$ are
\begin{eqnarray}
\mean{|\aCC_{lm}|^2} = \sum_{n} (U^l_{n})^2 P(k_{ln})
\;.\end{eqnarray}
Again, these aren't directly observable, but they contribute to the total CMB covariance, $\mean{|\hat a_{lm}|^2}$.

\Cfig{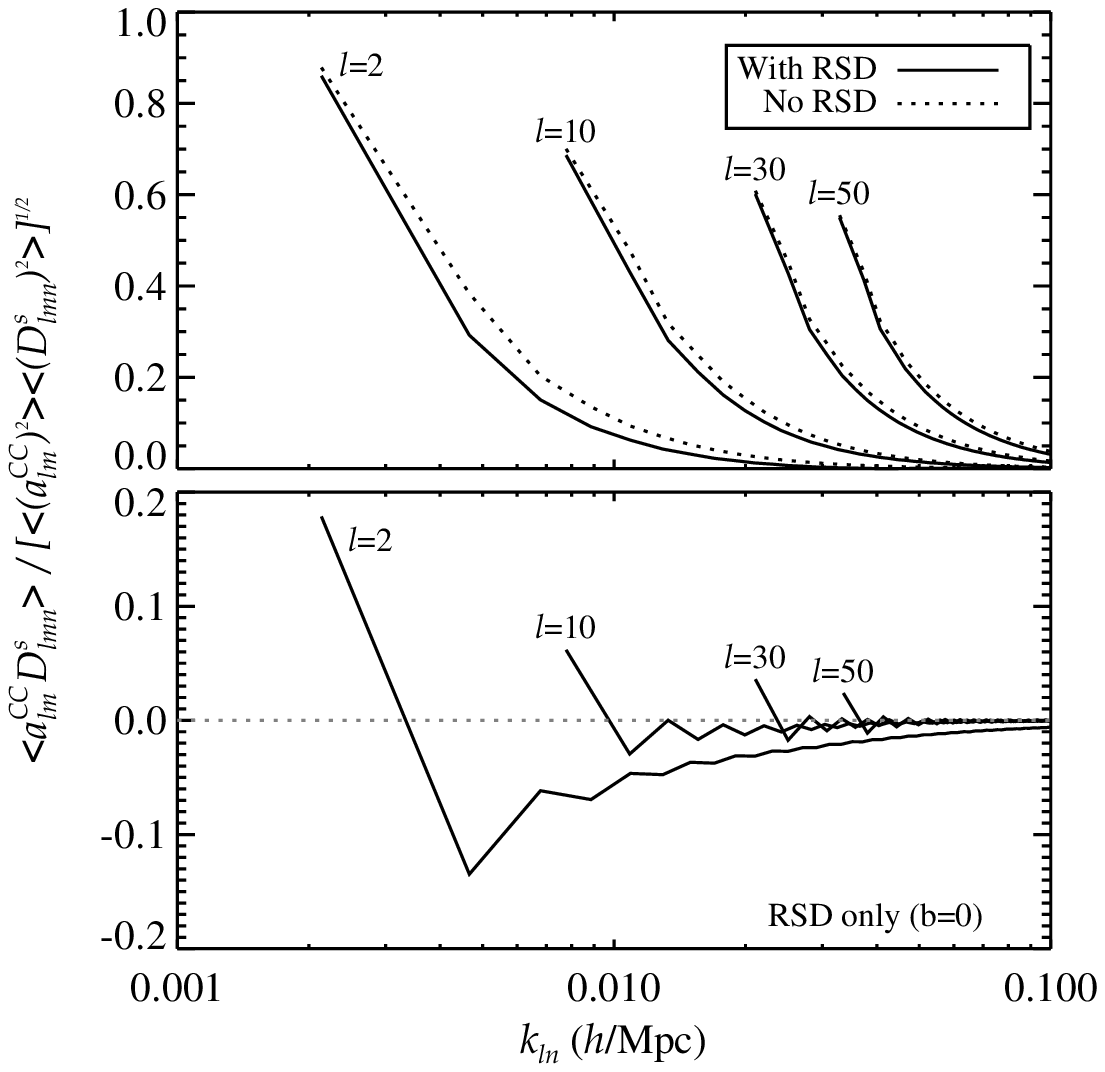}{Top: Pearson coefficients for CMB-galaxy cross-correlations.  The solid line includes the effects of RSDs while the dotted line ignores them. The $\aCC_{lm}$ are calculated using the ISW signal alone, and the ${\drhogal}^s_{lmn}$ neglect shot noise, therefore, these Pearson coefficients are theoretical upper limits for the signals alone.  Bottom: Pearson coefficient computed using the RSD term only ($b=0$).  The curve oscillations are \emph{not} due to numerical noise - they arise from the $V^l_{nn'}$ matrix.}

\Cfig{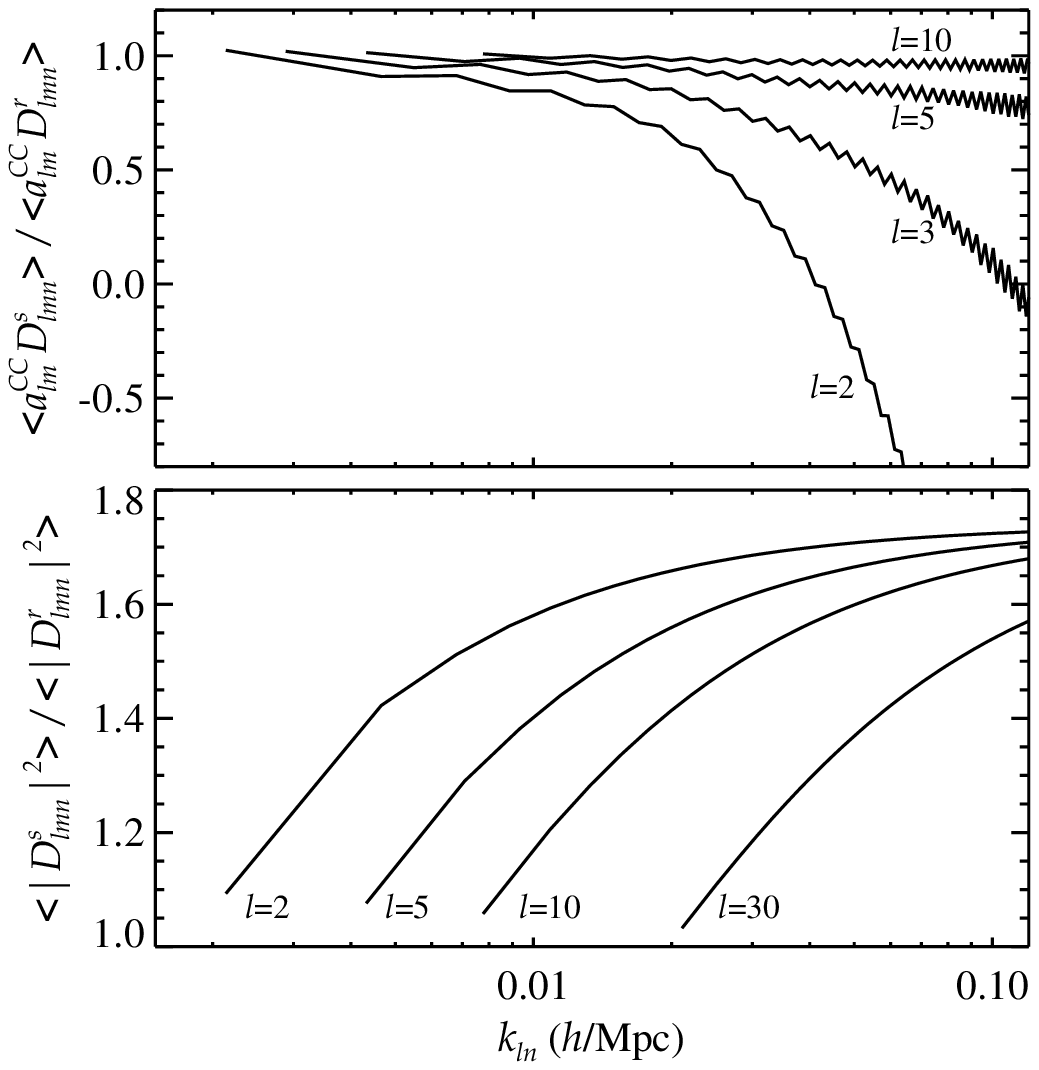}{The change in CMB-galaxy cross-correlations (top) and galaxy auro-correlations (bottom) due to RSDs.  ${\drhogal}^r_{lmn}$ is the same as ${\drhogal}^s_{lmn}$ defined in \refeq{gal_mode_simple} but with the RSD term removed.  The change in the cross-correlation is typically small, except on the largest angular scales where it can be significantly negative.  The curve oscillations are \emph{not} due to numerical noise - they arise from the $V^l_{nn'}$ matrix.}



It is interesting to look at how the presence of RSDs affects the correlation between the galaxy density and the CMB anisotropy.  
What we need are the Pearson correlation coefficients, defined as
\begin{equation}
R_{XY} \equiv \frac{\mean{XY}}{\sqrt{\mean{X^2}\mean{Y^2}}}
\end{equation}
for two random variables, $X$ and $Y$.  Two modes are fully correlated (contain completely redundant information) when their Pearson coefficient equals $\pm 1$, and they are completely independent if the Pearson coefficient is 0.  
We plot CMB-galaxy Pearson coefficients in the upper panel of \reffig{pearson_multi3}.  Note that the ${\drhogal}^s_{lmn}$ and $\aCC_{lm}$ plotted do not include noise or contributions from the non-ISW CMB.  These Pearson coefficients are therefore a theoretical upper limit -- 
the measurable quantities $\hat{\drhogal}^s_{lmn}$ and $\hat a_{lm}$ can only be less correlated.  

\reffig{pearson_multi3} shows that the inclusion of RSDs reduces the CMB-galaxy Pearson coefficients by a small amount for our modes of interest.  We highlight two reasons for this decrease in \reffig{rsd_correlation_effect}.  On small angular scales, it is primarily because RSDs increase the observed galaxy auto-correlations, while the CMB cross-correlations are left relatively unchanged by the RSD term in ${\drhogal}^s_{lmn}$.  Therefore, the effect on the Pearson coefficient calculation is that RSDs increase the denominator more than the numerator.  On the largest angular scales, CMB-galaxy correlations are also reduced because the ISW can correlate negatively with the RSD part of galaxy clustering.  To see why this happens, note the negative side-lobes of the $V^l_{nn'}$ matrix in \reffig{matrix_sections}, which are due to mode-mixing.  The $U^l_n$ matrix for the ISW is always positive, and when contracted with $V^l_{nn'}$, the negative contributions to the sum can dominate.  Hence the reduction in CMB-galaxy cross correlations is partially a geometric effect of the RSDs on large angular scales.

To further demonstrate the lack of correlation between the ISW and RSD signals, we can compute the Pearson coefficients for CMB-galaxy correlations using only the ``distortions'' -- the observed galaxy density fluctuations arising only from the RSD term in \refeq{drhogal}.  This computation is done by setting $b(0)=0$.  Thus we are imagining a galaxy density field with no real-space clustering but with observed redshift-space clustering due to peculiar velocities arising from the gravitational potentials.  It is clearly an unphysical situation; nevertheless, this Pearson coefficient allows us to understand where information about structure growth is coming from.  We want to discard the real-space galaxy clustering term, which has a very high signal-to-noise, but which is insensitive to $f\sigma_8(z)$.  We see in the lower panel of \reffig{pearson_multi3} that correlations between the ISW and RSDs do exist via the ``$VU$'' term in \refeq{CMB-gal-covar}, but they are at most 20\% on the largest angular and radial scales.  As before, this is a theoretical upper limit excluding noise and the early Universe CMB anisotropy, which can only reduce the correlations.  This low correlation, even in the absence of noise, occurs because the ISW and RSD signals depend on rather orthogonal combinations of the matter density modes under GR.  Since RSDs directly probe the time-like potential $\psi$ while the ISW is sourced by $\dot\psi-\dot\phi$, one might have hoped to use the two signals to test GR by measuring anisotropic stress on a mode-by-mode basis.  The low correlation between the signals implies that such an approach would not be feasible, but it does not prevent statistical tests of GR.

We further find that the presence of RSDs does not affect the signal-to-noise ratio for an ISW detection.  To compute the signal-to-noise, we first define $\hat C^{gT}_{ln}\equiv\mean{\hat{\drhogal}^s_{lmn}\hat a^*_{lm}}$.
The covariance of $\hat C^{gT}_{ln}$ is
\begin{equation}
\Covar(\hat C^{gT}_{ln},\hat C^{gT}_{ln'}) = \frac{\hat C^{gT}_{ln}\hat C^{gT}_{ln'}+\mean{|\hat a_{lm}|^2}\mean{\hat{\drhogal}^s_{lmn}\hat{\drhogal}^{s*}_{lmn'}}}{2l+1}
\end{equation}
and the total signal-to-noise is given by
\begin{equation}
\left(\frac{S}{N}\right)^2_{} = \sum_l\sum_{nn'}\hat C^{gT}_{ln} \hat C^{gT}_{ln'} \Covar^{-1}(\hat C^{gT}_{ln},\hat C^{gT}_{ln'})
\end{equation}
For our fiducial galaxy survey and Planck, we find that the total ISW signal-to-noise ratio is 4.25, and this value is negligibly affected by including or excluding RSDs.  \citet{afshordi_2004} calculates that an ideal survey extending to a redshift of 2 to 3 could detect the ISW at about 7.5$\sigma$.  The limiting factor for our signal-to-noise ratio is the cutoff of the galaxy survey at $z_{\max}=0.6$, and our result is consistent with Afshordi's findings at the lower redshift.

\section{Forecast for Combined Measurements of the Growth Rate} \labsec{forecast}



In this section, we forecast our ability to use galaxy and CMB maps to constrain $f\sigma_8$ -- the normalization of the linear velocity power spectrum -- as a function of redshift. Under GR, $f(z)\sigma_8(z)\approx \sigma_8(z)\Omega_M(z)^{\gamma(z)}$, where $\gamma(z)$ is nearly constant and weakly dependent on the dark energy equation of state \citep{Linder:2005in,Polarski:2008fk,peebles_1980,lahav_lilje_etal_1991,wang_steinhardt_1998}.  Instead of focusing on dark matter and dark energy parameters, we forecast direct constraints on $f\sigma_8$ itself in a few redshift bins.  Recall from \refsec{growth-compare} that $f\sigma_8$ is mainly probed by the ISW and by the RSD term in the redshift-space galaxy density.  The real-space galaxy clustering also constrains $b(z)\sigma_8(z)$; however we are less interested in this function, which is a constant in the case of CGC.

In general, for a set of cosmological parameters $p_\alpha$ that we wish to constrain, the Fisher matrix of the parameters is
\begin{equation}
F_{\alpha\beta} = \frac{1}{2} {\rm Tr}\left[ \Cov^{-1}\partder{\Cov}{p_\alpha}\Cov^{-1}\partder{\Cov}{p_\beta} \right]
\end{equation}
where $\Cov$ is the full covariance matrix for all observables.  The forecasted errors on the $p_\alpha$ are $\sigma_{\rm fix}(p_\alpha)=F_{\alpha\alpha}^{-1/2}$ if all other parameters are held fixed or $\sigma_{\rm marg}(p_\alpha)=(F^{-1})_{\alpha\alpha}^{1/2}$ if we marginalize over the other parameters.

Our observables are the $\hat a_{lm}$, the spherical harmonic coefficients of the CMB anisotropy, and the $\hat {\drhogal}^s_{lmn}$, the SFB coefficients of the galaxy number density.  We combine these into a joint data set, $\hat\joint_{lmn}$:
\begin{equation}
\hat\joint_{lmn} \equiv \left\{
\begin{array}{cc}
  \hat a_{lm} & (n=0) \\
  \hat{\drhogal}^s_{lmn} & (n>0) \\ 
\end{array} \right.
\;.\end{equation}
The joint covariance matrices are $\Cov_{ll'mm'nn'}=\mean{\hat\joint_{lmn}\hat\joint^*_{l'm'n'}}$.  Due to the symmetry of our fiducial survey and the linearity of our cosmological model, the $\Cov$ will be independent of $m$ and block-diagonal in $m$ and $l$.  Therefore, we need only calculate the much simpler submatrices, $\Cov^l_{nn'}\equiv\mean{\hat\joint_{lmn}\hat\joint^*_{lmn'}}$, and we find that
\begin{eqnarray}
F_{\alpha\beta} &=& \frac{1}{2} \sum_l (2l+1) {\rm Tr}\left[ (\Cov^l)^{-1}\partder{\Cov^l}{p_\alpha}(\Cov^l)^{-1}\partder{\Cov^l}{p_\beta} \right] \nonumber \\
 &=& \frac{1}{2} \sum_l (2l+1) \nonumber \\
 & & \moreeq \sum_{nn' \atop n''n'''} (\Cov^l)^{-1}_{nn'}\partder{\Cov^l_{n'n''}}{p_\alpha}(\Cov^l)^{-1}_{n''n'''}\partder{\Cov^l_{n'''n}}{p_\beta}
\end{eqnarray}


\Cfig{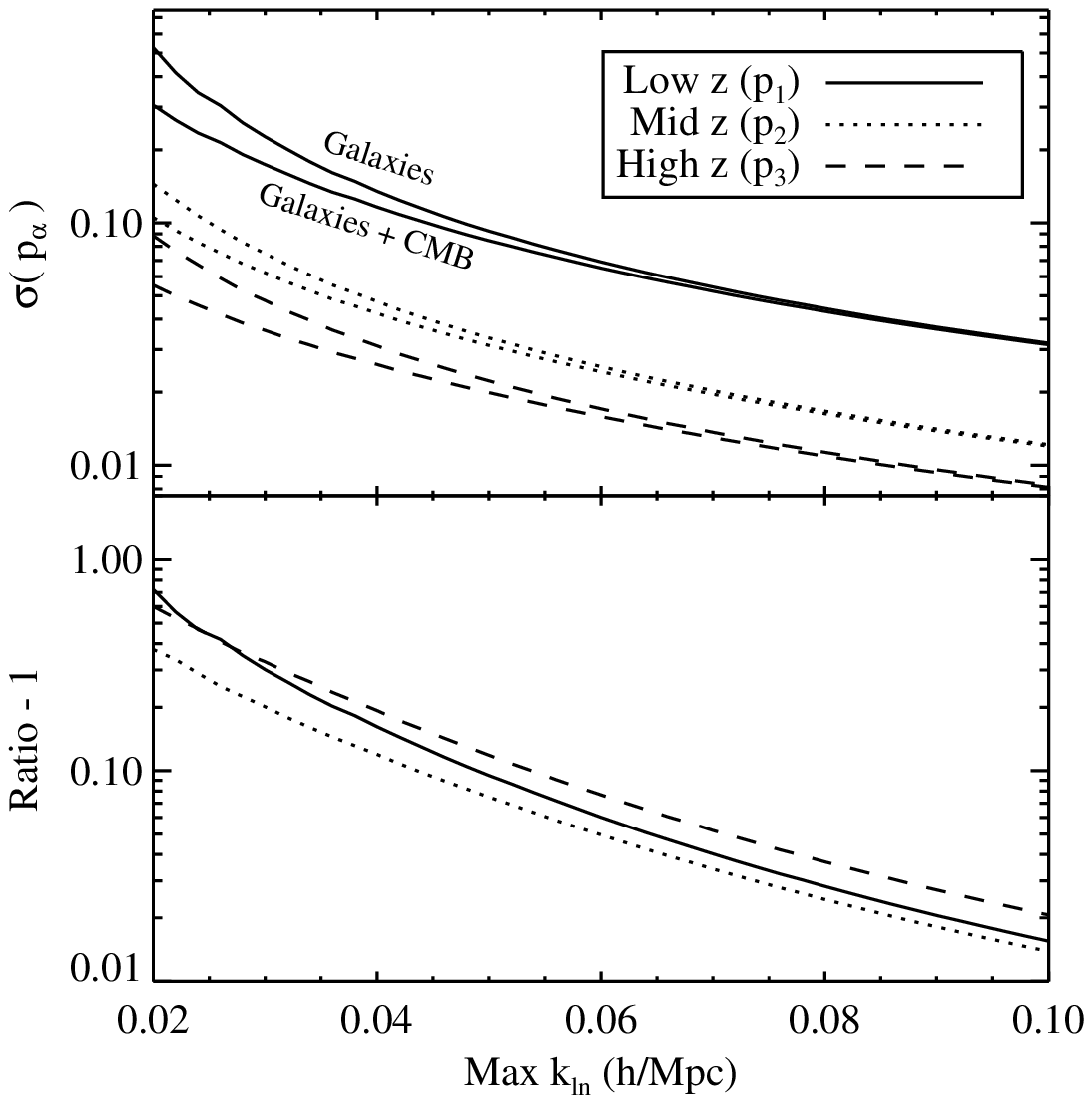}{Top panel: Forecasted error bars on $p_\alpha$, the offsets to $f\sigma_8$, as a function of maximum $k$. The solid, dotted and dashed curves are for the low, medium and high redshift bins, respectively.  For each pair of curves, the upper curve uses galaxies alone while the lower curve includes CMB-galaxy cross-correlations.  The constraints on each $z$ bin have been marginalized over the other two bins and the normalization $b\sigma_8(z=0)$.  Bottom panel: For each pair of curves in the top panel, the ratio (upper over lower) minus one.}

The $p_\alpha$ for which we forecast constraints will be offsets to $f\sigma_8$ -- relative to \LCDM -- in three redshift bins.  We divide our redshift range into three bins of equal width in $\log(1+z)$.  The bin boundaries are $z_0=0, z_1=0.170, z_2=0.368, z_3=0.6$.  We start with a fiducial $f\sigma_8(z)$ computed using \LCDM and allow this function to have a piece-wise constant offset in each bin:
\begin{equation} \label{eq:pdef}
f\sigma_8(z) = f\sigma_8(z;\LCDMeq) + p_\alpha \hspace{.5cm} {(z_{\alpha-1}<z \le z_{\alpha})}
\end{equation}
for $\alpha$=1,2, or 3.  We do not specify $f\sigma_8(z)$ beyond the survey boundary ($z>z_{\max}$), but it is constrained by the normalization of the CMB.  When we modify $f\sigma_8(z)$, we then compute the linear growth $G(z)$ by integrating $fG=\partder{G}{\ln a}$, normalizing so that $\sigma_8(0)$ remains fixed.  Additionally, we include the normalization of the present galaxy power spectrum as a nuisance parameter: $p_0 = b\sigma_8(0)$.
Although RSDs and the ISW have some sensitivity to other cosmological parameters, such parameters are left fixed in our forecast since we expect them to be tightly constrained by other observables.  For instance, the ISW is sensitive to $\Omega_mh^2$, but this parameter is well-measured by the first acoustic peak of the CMB.  We compute $\partder{\Cov}{p_\alpha}$ using two-sided derivatives with steps of $\Delta p_\alpha=$0.01 or 0.005 and find little difference in the results.

\reffig{varyk} shows our forecasted constraints on the $p_\alpha$, with and without CMB information, as a function of our cutoff scale for the observed galaxy density, $k_{\max}$.  The errors have been marginalized over the other parameters, including the normalization $b\sigma_8(0)$.  The bottom panel of \reffig{varyk} shows the extent to which using the ISW has improved the constraints on $f\sigma_8(z)$ relative to using RSDs alone.  The plot shows that for $k<0.05 h$/Mpc, ISW measurements would improve constraints in each redshift bin by more than 10\%.  When RSD measurements are available on smaller scales, the ISW does not significantly improve constraints on a scale-independent $f\sigma_8(z)$.  Our constraints for $k_{\max}=0.1 h$/Mpc are in good agreement with the results of \citep{white2009}, who use the flat-sky and distant observer approximations.  
To test the sensitivity of our results to boundary conditions, we tried recomputing the $f\sigma_8$ constraints using different values of $z_{\max}$ (0.6, 0.8 and 1.0) while keeping the definition for $p_\alpha$ in \refeq{pdef}.  We find that our results in \reffig{varyk} are reasonably robust, and summarize them in \refsec{ksamp}.


\section{Conclusions} \labsec{conclusions}

The integrated Sachs-Wolfe effect and redshift-space distortions are unique cosmological probes in that they are sensitive to the evolution of large-scale structure rather than simply the amount of structure at a given epoch.  In general, they probe different combinations of the linear gravitational potentials, $\phi$ and $\psi$; however, under GR, they both can be shown to measure $f\sigma_8(z)$, the amplitude of the peculiar velocity power spectrum.  We investigated whether the two probes were complementary or redundant when combined to constrain $f\sigma_8(z)$ using data from the Planck CMB experiment and a large galaxy survey with a redshift distribution similar to that expected for BOSS.  Our analysis used the spherical Fourier-Bessel expansion in order to account for important large-angle effects and to avoid discarding information via redshift-binning.  The SFB basis also provides an original insight into the effects of RSDs on CMB-galaxy cross-correlations, and it allows us to make a mode-by-mode comparison of the correlation between RSD and ISW measurements.

We find that the ISW and RSDs are mostly independent (uncorrelated) observables, being sensitive to mostly orthogonal combinations of the matter density field. In general, RSD depend on radial modes, while the ISW is more sensitive to angular fluctuations in the galaxy density.  We also find that the ISW, measured through CMB-galaxy cross-correlations, improves $f\sigma_8(z)$ constraints from RSDs by only 10\% even when the analysis is restricted to large physical scales ($k>0.05 h$/Mpc).  Thus, when future precision measurements of RSDs are available, the ISW will be more valuable as a probe of non-standard GR models and as a test of survey systematics and less valuable as a way to measure cosmological parameters in GR.

\section*{Acknowledgments}

Thanks to Alan Heavens, Antony Lewis, and Rita Tojeiro for their helpful insights.  Thanks to Cyril Pitrou and Kazuya Koyama for assistance with code debugging.  We also thank our anonymous referee for constructive criticisms which have improved this manuscript.  Calculations were done in part by modifying the \texttt{iCosmo} IDL package \citep{icosmo_2011}.  CS, RC and WP are funded by an STFC Rolling Grant.  WP also acknowledges funding from the European Research Council and the Leverhulme Trust.

\appendix

\section{$k$-space Sampling} \label{sec:ksamp}

\Cfig{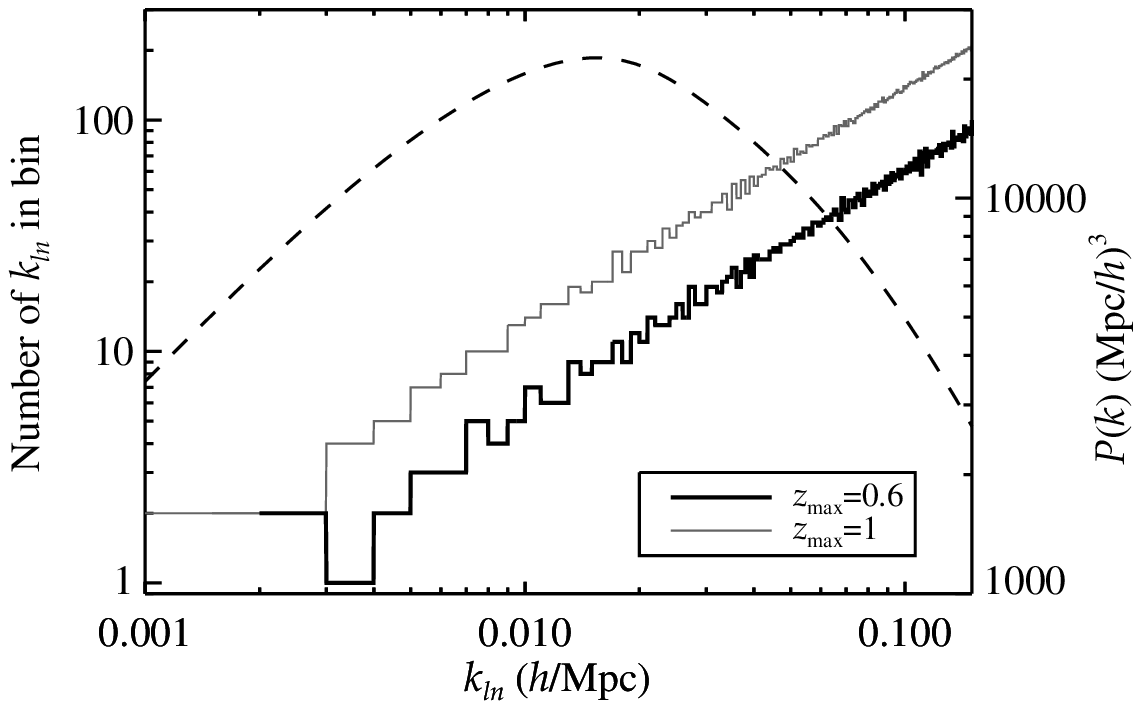}{Histogram showing how $P(k)$ is sampled by the discrete $k_{ln}$ of the SFB expansion.  The dark and light solid lines show the number of $k_{ln}$ in bins of $\Delta k=0.001 h$/Mpc for $z_{\max}=0.6$ and $z_{\max}=1$, respectively.  Both assume Neumann boundary conditions.  The linear matter power spectrum is shown by the dashed line.}

The discrete $k_{ln}$ needed for our SFB analysis are determined by the boundary conditions.  \reffig{k1_k06_histogram_Pk} illustrates the sampling in $k$-space corresponding to the Neumann boundary conditions in \refeq{boundary}.  Changing $z_{\max}$ changes the values of the $k_{ln}$, the number of $k_{ln}$ less than $k_{\max}$, and subsequently how $P(k)$ is sampled.  In particular, the minimum wavenumber is given by $k_{21}=3.34/r_{\max}$, which in our case is 0.00214 $h$/Mpc for $z_{\max}$=0.6 and 0.00142 $h$/Mpc for $z_{\max}$=1.  Since our Fisher matrix analysis is sensitive to $P(k)$, it will be somewhat sensitive to our boundary conditions.

We tested the sensitivity of the $f\sigma_8$ constraints in \reffig{varyk} to boundary conditions by varying the $z_{\max}$ which defines the SFB expansion.  We choose to compare the {\it unmarginalized} errors for different boundary conditions since increasing the survey size significantly improves constraints on $b\sigma_8$, which is partially degenerate with the $f\sigma_8$ constraints.  For $z_{\max}=0.8$ and $k_{\max}=0.1$, we find that the unmarginalized error bars on $f\sigma_8$ change negligibly in the two lower-$z$ bins, and they degrade by less than 2\% for the highest $z$ bin.  
For $z_{\max}=0.8$ and $k_{\max}=0.02$, the errors degrade by at most 10\% for the high-$z$ bin (with ISW).  The larger effect for the lower $k_{\max}$ reflects the fact that the lower $k$ are more sparsely sampled; therefore changes to the sampling of $P(k)$ have a greater impact on constraints relying on the lower $k$.  Pushing the boundary further to $z_{\max}=1$ results in negligible changes from $z_{\max}=0.8$.

We could have chosen to make $z_{\max}\gg 1$ so as to have the finest possible sampling in $k$-space; however, doing so adds considerable computational difficulty without much benefit.  In addition to increasing the number of wavenumbers, our galaxy survey would require a selection function that drops the galaxy density to zero above $z \sim 1$, resulting in very significant mode-mixing in the $\Phi$ and $V$ matrices in \refeq{gal_mode_simple}.

\section{Angular Correlations in Redshift Slices}  \label{sec:slice}

\Cfig{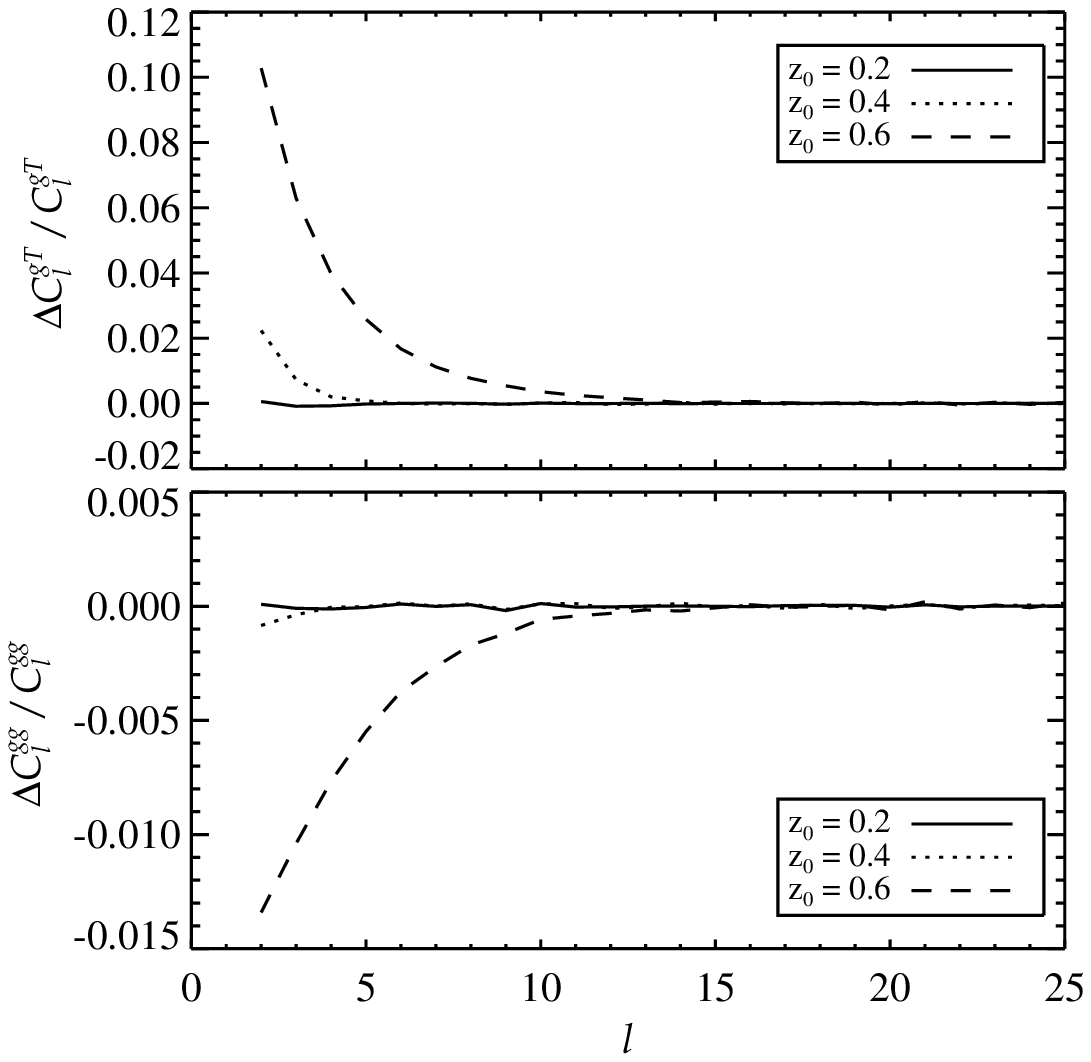}{The effects of boundary conditions on computations of CMB-galaxy angular cross-correlations (top) and galaxy auto-correlations (bottom) in redshift slices.  The denominators, $C_l^{gT}$ and $C_l^{TT}$, are computed by summing coefficient correlations in \refeq{resum_cross} and \refeq{resum_auto}, with the boundary set at $z_{\max}=1.0$.  $\Delta C_l$ is $C_l(z_{\max}=0.8)-C_l(z_{\max}=1.0)$.  The redshift slices are Gaussians centered at $z_0$ with $\sigma_z=0.02$.}

A more familiar method of computing correlations of galaxy number counts is to bin galaxies into redshift slices.  The 2D projected galaxy density in each slice can then be auto-correlated or cross-correlated with other slices or the CMB temperature.  In this section, we show how to convert our 3D SFB results into the more common 2D representation.

Start with the redshift-space galaxy density contrast expanded into SFB modes:
\begin{equation}
\drhogal(\vs) = \sum_{lmn} c_{ln}\drhogal^s_{lmn}j_l(k_{ln}s)Y_{lm}(\skypos)
\end{equation}
The projected density contrast in a redshift slice is
\begin{equation}
\drhogal_2(\skypos) = \int_0^{s_{\max}}\!\!\!\!\!\! \dint{}{s} w(s) \drhogal(\vs)
\end{equation}
where $w(s)$ is a radial window function defined in redshift-space.
The spherical harmonic transform of $\drhogal_2(\skypos)$ is
\begin{eqnarray}
\drhogal^s_{lm} &=& \int_{4\pi} d^2\theta\; Y_{lm}^*(\skypos)\drhogal_2(\skypos) \nonumber \\
&=& \sum_n W^s_{ln} \drhogal^s_{lmn}
\end{eqnarray}
with
\begin{equation}
W^s_{ln}\equiv c_{ln} \int_0^{s_{\max}}\!\!\!\!\!\! \dint{}{s} w(s) j_l(k_{ln}s)
\;.\end{equation}
Correlating these projected density modes with the CMB temperature shows that the 2D/3D CMB-galaxy correlations can be summed accordingly to obtain their fully 2D angular correlation in the slice:
\begin{equation} \labeq{resum_cross}
C_l^{gT}\equiv\mean{\aCC_{lm}{\drhogal}^{s*}_{lm}} = \sum_n W^s_{ln} \mean{\aCC_{lm}{\drhogal}^{s*}_{lmn}}
\;. \end{equation}
As before, the symmetry of our fiducial survey ensures independence from $m$ and delta functions in $l$ and $m$.
Similarly, the autocorrelation of galaxy clustering can be written
\begin{equation} \labeq{resum_auto}
C_l^{gg}\equiv\mean{{\drhogal}^s_{lm}{\drhogal}^{s*}_{lm}} = \sum_{nn'} W^s_{ln}W^s_{ln'} \mean{{\drhogal}^s_{lmn}{\drhogal}^{s*}_{lmn'}}
\;.\end{equation}


Computing $C_l^{gg}$ and $C_l^{gT}$ using the above sums is a bit challenging since since their accuracy depends on the boundary conditions (see \refsec{ksamp}) and on the maximum $k$.  The expressions should converge for large enough $s_{\max}$ and $k$, but the computation becomes intensive due to mode-mixing and the increasing number of discrete modes.  More manageable expressions for computing galaxy correlations in redshift slices (with RSDs) are given by \citet{padmanabhan07} and \citet{rassat_2009}.

To demonstrate the limitations of our approach, we compute $C_l^{gg}$ and $C_l^{gT}$ for several redshift slices and boundary conditions.  We use Gaussian slices with $\mean{z}=z_0$ and $\sigma_z=0.02$ so that the window function  $w\propto\exp[-\frac{1}{2}(z-z_0)^2/\sigma_z^2]$.  \reffig{compare_z8toz1_slice_dz02} shows how our angular correlation calculations change when we change boundary conditions from $z_{\max}=0.8$ to $z_{\max}=1.0$.  We see that both the galaxy count auto-correlations and the CMB-galaxy cross-correlations are affected by several percent on the largest angular scales.  This is because the characteristic scale of a slice, $k\approx l/r(z_0)$, lies in a poorly sampled region of the $k_{ln}$ for small $l$ (see \reffig{k1_k06_histogram_Pk}).  For example, with $l=2$ and $z_0=0.6$, we have $l/r(z_0)=0.00128\,h$/Mpc.
We reiterate that these boundary effects have only a small impact on our main result, which is the forecast in \refsec{forecast}.


\label{lastpage}

\newpage

\bibliography{ISW_RSD} 
\bibliographystyle{mn2e}

\end{document}